\documentclass{naturefig}
\usepackage{amsmath}
\usepackage{amssymb}
\usepackage{graphicx}
\usepackage{color}

\newcommand{\beginsupplement}{%
        \setcounter{table}{0}
        \setcounter{figure}{0}
     }

\providecommand{\B}{\rm B}
\providecommand{\F}{\rm F}
\providecommand{\BB}{\rm BB}
\providecommand{\FB}{\rm FB}
\providecommand{\FS}{\rm FS}

\newcommand{\bra}[1]{\left\langle #1 \right\rvert}
\newcommand{\ket}[1]{\left\lvert #1 \right\rangle}
\newcommand{\xv}[1]{\left\langle #1 \right\rangle}

\title{Observation of coherent quench dynamics in a metallic many-body
  state of fermionic atoms}

\author{Sebastian Will$^{1,2}$, Deepak Iyer$^{3}$, Marcos
  Rigol$^{3}$}

\begin{document}

\date{\today}
\maketitle
\begin{affiliations}
\item Department of Physics, Massachusetts Institute of Technology,
  Cambridge, MA 02139, USA
\item Institut f\"ur Physik, Johannes Gutenberg-Universit\"at, 55099
  Mainz, Germany
\item Department of Physics, The Pennsylvania State University,
  University Park, PA 16802, USA
\end{affiliations}


\begin{abstract}
Quantum simulation with ultracold atoms has become a powerful technique to gain insight into interacting many-body systems. In particular, the possibility to study nonequilibrium dynamics offers a unique pathway to understand correlations and excitations in strongly interacting quantum matter. So far, coherent nonequilibrium dynamics has exclusively been observed in ultracold many-body systems of bosonic atoms. Here we report on the observation of coherent quench dynamics of fermionic atoms. A metallic state of ultracold spin-polarised fermions is prepared along with a Bose-Einstein condensate in a shallow three-dimensional optical lattice. After a quench that suppresses tunnelling between lattice sites for both the fermions and the bosons, we observe long-lived coherent oscillations in the fermionic momentum distribution, with a period that is determined solely by the Fermi-Bose interaction energy. Our results show that coherent quench dynamics can serve as a sensitive probe for correlations in delocalised fermionic quantum states and for quantum metrology. 
\end{abstract}


\baselineskip24pt


\newpage
\section*{Introduction}

The investigation of nonequilibrium dynamics in interacting
  quantum many-body systems has emerged as a major research direction
  in the field of ultracold atoms.  It provides unique insight into
  quantum states, their excitation
  spectra\cite{lewenstein07,bloch_dalibard_review_08,cazalilla_citro_review_11},
  and thermalisation processes\cite{rigol08,polkovnikov11}. Time
  evolution far from equilibrium has primarily been studied using
  purely bosonic systems, allowing the observation of coherent quench
  dynamics\cite{greiner02,sebby07,will10,meinert14} and
  equilibration\cite{kinoshita06, trotzky12, gring12} in isolated
  setups. Close to equilibrium, various driving protocols have
    been devised to use dynamics as a means for obtaining information
    about the excitation spectra of many-body phases in optical lattices\cite{iucci_cazalilla_06,kollath06_iucci_06a,tokuno_giamarchi_11,he_brown_14}. 
    Centre of mass oscillations in a mixture of fermionic and
    bosonic superfluids have been used to measure the coupling between
    the two superfluids\cite{barbut14}. In purely fermionic systems,
  nonequilibrium dynamics has been explored in transport measurements
  that allowed for a semi-classical theoretical
  description\cite{Schneider12}. However, so far, the observation of
  coherent nonequilibrium quantum dynamics for fermions has remained
  elusive.

At ultracold temperatures, quantum statistics dominates
 and gives rise to distinctive many-body ground states
for bosonic and fermio\-nic systems\cite{feynman_book_98}.
Noninteracting bosons collectively condense into the single-particle
state of lowest energy, forming a Bose-Einstein condensate (BEC). Fermions, on the other hand,
obey the Pauli exclusion principle, which limits the occupation of
single-particle states to a maximum of one fermion. Therefore,
fermions fill the lowest energy single-particle states from bottom up
and form a Fermi sea. When placed in a periodic lattice potential with
$M$ sites, the wavefunction of the BEC can be written as
  the product\cite{zwerger03} $\ket{\Psi_{\rm
    BEC}}=\prod_{j=1}^M\ket{\alpha_j}$ of identical coherent states
$\ket{\alpha_j}=e^{-|\alpha|^2/2}
e^{\alpha\,\hat{a}^{\dag}_j}\ket{0}$, where $|\alpha|^2$ is the mean
occupation per lattice site, and $\hat{a}^{\dag}_j$ the bosonic
creation operator at site $j$. On the other hand, the
wavefunction of an ideal Fermi gas of $N$ identical fermions can
be expressed by the product $\ket{\Psi_{\rm
    FS}}=\prod_{E_{\mathbf{k}}\leq E_{\F}}|\mathbf{k}\rangle$ of the
$N$ quasi-momentum eigenstates
$|\mathbf{k}\rangle=M^{-1/2}\sum_{j=1}^M
e^{i\mathbf{k}\cdot\mathbf{r}_{j}} \hat{c}^{\dag}_{j}\ket{0}$ with
energy eigenvalues $E_{\bf k}$ smaller than the Fermi energy $E_{\rm
  F}$.  Here, $\hat{c}^{\dag}_j$ denotes the fermionic creation
operator, and $\mathbf{r}_{j}$ the position of site $j$. As long as
the fermions do not completely fill up a lattice band, $\ket{\Psi_{\rm
    FS}}$ represents a metallic state.
    
The distinct ground state properties of bosons and fermions have
direct implications for their respective many-body quantum
dynamics. For the case of bosons, coherent quench dynamics was
experimentally studied by preparing an atomic BEC in a shallow optical
lattice and taking it out of equilibrium by a sudden quench to a deep
lattice\cite{greiner02,sebby07,will10}. The rapid suppression of
tunnelling and the enhanced interactions between the atoms gave rise
to characteristic collapses and revivals of the bosonic matter wave
interference pattern, whose periodicity is determined by the strength
of the on-site interaction $U^{\BB}$. In homogeneous lattice
potentials, this phenomenon can be understood from the dynamics of a
single lattice site: The time evolution of the many-body state is
governed by the operator $e^{-i\hat{H}t/\hbar}=\prod_{j=1}^M
e^{-i\hat{H}_jt/\hbar}$ with
$\hat{H}_j=U^{\BB}\hat{n}_j(\hat{n}_j-1)/2$ being the on-site
interaction term of the Bose-Hubbard Hamiltonian, where
$\hat{n}_j=\hat{a}^{\dag}_j\hat{a}_j$ counts the number of bosons at
site $j$. Consequently, the dynamics of the entire system,
$e^{-i\hat{H}t/\hbar}\ket{\Psi_{\rm BEC}}= \prod_{j=1}^M
e^{-i\hat{H}_jt/\hbar}\ket{\alpha_j}$, is comprised of a product of
identical dynamics at each lattice site.

In this work, we are concerned with the dynamics of a delocalised
many-body state of fermions for which, even in a homogeneous lattice,
an effective single-site description is not
possible. Specifically, we consider a shallow optical lattice
  that is simultaneously loaded with a metallic state of
  spin-polarised fermionic atoms and an atomic BEC, as schematically
  shown in Fig.~\ref{fig:1}{a}. Initially, the interactions between
fermions and bosons are weak, while the large kinetic energy
  dominates. Therefore, we approximate the quantum state of
  this hybrid Fermi-Bose system by the direct product
  $\ket{\Psi_\text{FS}}\otimes\ket{\Psi_\text{BEC}}$. When the system
is quenched by a rapid increase of lattice depth, tunnelling between
lattice sites is suppressed both for fermions and bosons and
interparticle interactions dominate. Interactions among the bosonic
component give rise to typical collapse and revival dynamics that has
been analysed previously\cite{will11}. It is the key finding of the
present work that the fermionic component also undergoes coherent
dynamics. Although the fermions do not interact among themselves,
their interaction with the bosons drives the dynamics of the
quenched metallic state. Similar to the purely bosonic case, the time
evolution operator $e^{-i\hat{H}t/\hbar}$ factorises into
$\prod_{j=1}^M e^{-i\hat{H}^{\FB}_jt/\hbar}$ with
$\hat{H}^{\FB}_j=U^{\FB}\hat{n}_j\hat{m}_j$ being the on-site
interaction term of the Fermi-Bose Hubbard Hamiltonian\cite{albus03},
where $\hat{m}_j=\hat{c}^{\dag}_j\hat{c}^{}_j$ counts the number of
fermions at site $j$ and $U^{\FB}$ is the on-site Fermi-Bose
interaction energy.  However, due to its delocalised nature, the
initial metallic state $\ket{\Psi_\text{FS}}$ does not factorise into
a product of on-site wavefunctions. This is crucial for coherent
fermionic dynamics to occur, the time scale of which is given by
  $h/U^{\rm BF}$. We discuss the role of fermionic and bosonic number
  fluctuations in the quench dynamics and introduce the visibility of the fermionic momentum distribution as a suitable observable. In the
  experiment, we study the dynamics for various strengths of the
  Fermi-Bose interaction and reveal their spectral properties through
  Fourier analyses. As a result of fermionic quantum statistics, the
  spectra exclusively reveal the Fermi-Bose interaction energy
  $U^{\FB}$ with high resolution. Coherent quench dynamics can
  therefore be used as a sensitive probe for correlations and complex
  interaction effects in hybrid many-body quantum systems.


\begin{figure}
  \begin{center}
    \includegraphics[width=0.54\columnwidth]{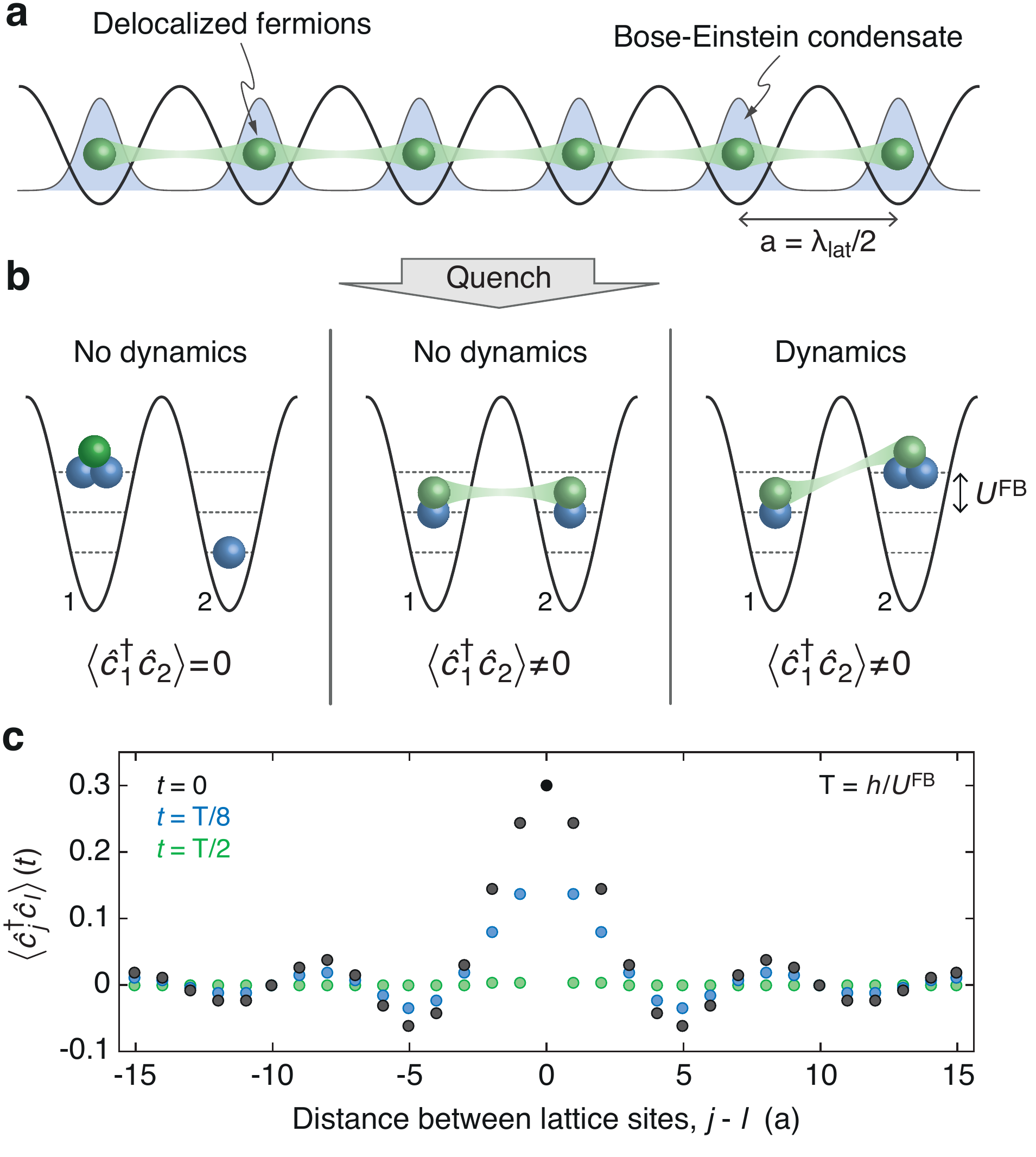}
  \end{center}
  \caption{\label{fig:1} \textbf{Emergence of coherent quench dynamics
      in a metallic many-body state.} \textbf{a,} A Fermi-Bose quantum
    gas mixture is loaded into a shallow optical lattice. The
    spin-polarised fermions form a metallic many-body quantum state,
    delocalised across the lattice (green spheres). The bosons form a
    BEC (blue background). \textbf{b,} A rapid increase of lattice
    depth quenches the system. The on-site interactions between
    fermions and bosons, $U^{\rm FB}$, drive coherent dynamics in the
    fermionic momentum distribution. The origin of the dynamics is
    illustrated by considering three sample cases in a two-site
    system.  If the system features a localised fermion (left) or
    bosonic Fock states with an equal number of bosons on each site
    (middle), no dynamics occurs. Dynamics occurs if the fermion is
    delocalised, $\langle\hat{c}_1^\dagger\hat{c}^{}_2\rangle\neq 0$,
    and the number of bosons is different on each site
    (right). \textbf{c,} Fermionic single-particle correlations in a
    1D lattice at times $t=0$, $h/(8U^{\rm FB})$, and $h/(2U^{\rm
      FB})$ after the quench. The fermionic filling is chosen to be
    $\bar{m} =0.3$ and stays constant as a function of time (black
    point at $j-l=0$).}
\end{figure}


\section*{Results}

\subsection{Quench dynamics in a two-site system.}

In order to illustrate the emergence of coherent quench dynamics, we consider an
elementary setup with two lattice sites\cite{Schachenmayer11} (labeled 1 and 2, 
spaced by distance $a$), occupied by a single fermion and multiple bosons (see
Fig.~\ref{fig:1}{b}). For finite tunnelling and vanishing interactions,
$U^{\FB}=0$, the fermionic ground state has the form
$(\hat{c}^{\dag}_{1}+\hat{c}^{\dag}_{2})\ket{0}/\sqrt{2}$,
corresponding to the fermion being in the $k=0$ quasi-momentum
eigenstate. After a quench in which interactions between fermions 
and bosons are turned on and tunnelling amplitudes for fermions and bosons 
are set to zero, the site occupations
$\langle\hat{c}^{\dag}_{1}\hat{c}^{}_{1}\rangle =
\langle\hat{c}^{\dag}_{2}\hat{c}^{}_{2}\rangle =1/2$ remain constant
due to the absence of tunnelling.  The off-diagonal correlations of
the single-particle density matrix, however, evolve in time:
$\langle\hat{c}^{\dag}_{1}\hat{c}_{2}\rangle = e^{-i U^{\rm
    FB}(n_{1}-n_{2}) t/\hbar}/2$, where $n_1$ ($n_2$) is the number of
bosons on site 1 (2). Consequently, the fermionic quasi-momentum
distribution, $n (k) \equiv M^{-1} \sum_{j,l}e^{ik a (j-l)}\langle
\hat{c}^{\dag}_{j}\hat{c}^{}_{l}\rangle$, undergoes dynamics. The
evolution for the two allowed quasi-momentum eigenstates $k=0$ and $k=\pi/a$
reads $(1\pm\cos[U^{\FB}(n_{1}-n_{2})t/\hbar])/2$, respectively, and
indicates that the fermion oscillates between the $k=0$ and $k=\pi/a$
states with a period $T\propto h/U^{\FB}$. In contrast, no dynamics
occurs if the fermion initially occupies a localised state,
$\hat{c}^{\dag}_{1}\ket{0}$ or $\hat{c}^{\dag}_{2}\ket{0}$, since the
off-diagonal correlations vanish, or if the number of bosons is
identical on both sites ($n_{1}=n_{2}$). For $n(k)$ to evolve with
time, delocalised fermions and spatially varying bosonic occupancies
are required. In the experiment, the latter is provided by the quantum
fluctuations of the on-site occupation that are characteristic for a
BEC. Since this specific example of a two-site model only contains one fermion, quantum statistics does not play a role in the dynamics of 
$n(k)$. However, quantum statistics plays an essential role in the dynamics of the 
quasi-momentum distribution when, as in the experiments, many fermions are 
present.
 
\subsection{Fermionic quench dynamics in a lattice system.}
 
To characterise the quench dynamics in a setup with many lattice
sites, it is convenient to study the visibility of the fermionic 
quasi-momentum distribution. We define it as the ratio between 
the number of fermions with quasi-momenta in the interval $[-k_0,k_0]$ 
(see Fig.~\ref{fig:2}{a}) and the total fermion number $N$, 
$\mathcal{V}_{\rm F}\equiv\frac{1}{N}\int_{-k_0}^{k_0}{\rm d}k_x\,n(k_x)$. 
For the case of a 1D lattice and an initial state
$\ket{\Psi_\text{FS}}\otimes\ket{\Psi_\text{BEC}}$, the time evolution
of the visibility after the quench can be calculated analytically in
the thermodynamic limit (see Methods). Assuming $k_0 \leq k_{\rm F}$, we obtain
\begin{equation}\label{eq:vis-notrap}
  \mathcal{V}_{\rm F}(t) = \frac{k_{0}}{k_{\rm lat}} + \left[
    \frac{k_{0}}{k_{\F}} -\frac{k_{0}}{k_{\rm lat}}
  \right]e^{2|\alpha|^2 \left[ \cos \left( U^{\FB}t / \hbar \right) -1 \right]},
\end{equation}
where $k_{\rm F}$ is the Fermi quasi-momentum, $k_{\rm lat} = \pi/a$ is the
quasi-momentum corresponding to the edge of the Brillouin zone and $a$
is the lattice spacing. Analogous to the two-site case, the
periodicity of the oscillation is determined by $T=h/U^{\FB}$, and the
coherent dynamics originates from the presence of off-diagonal single
particle correlations and on-site occupancy fluctuations of the
BEC. Figure \ref{fig:1}{c} illustrates the time evolution of the
off-diagonal correlations $\langle
\hat{c}^{\dag}_{j}\hat{c}^{}_{l}\rangle(t)=\bar{m}
\sin[\pi\bar{m}(j-l)]\exp[2|\alpha|^{2}\{\cos(U^{\rm FB}
t/\hbar)-1)\}]/(j-l)$, where $\bar{m} = k_{\rm F} / k_{\rm lat}$ is
the fermionic filling. The theoretical analysis can be extended to three-dimensional (3D)
lattices, where $n(k_x,t)$ is taken to be the projection of the full
fermionic quasi-momentum distribution $n(k_x,k_y,k_z,t)$ onto one 
dimension, $n(k_x,t)=\int_{-k_{\rm lat}}^{k_{\rm lat}} {\rm d}k_y \int_{-k_{\rm
    lat}}^{k_{\rm lat}} {\rm d}k_z \, n(k_x,k_y,k_z,t)$.  The results
are qualitatively similar to Eq.~\eqref{eq:vis-notrap} (see Methods).

\subsection{Experimental sequence and observation of quench dynamics.}

\begin{figure}
  \begin{center}
    \includegraphics[width=0.44\columnwidth]{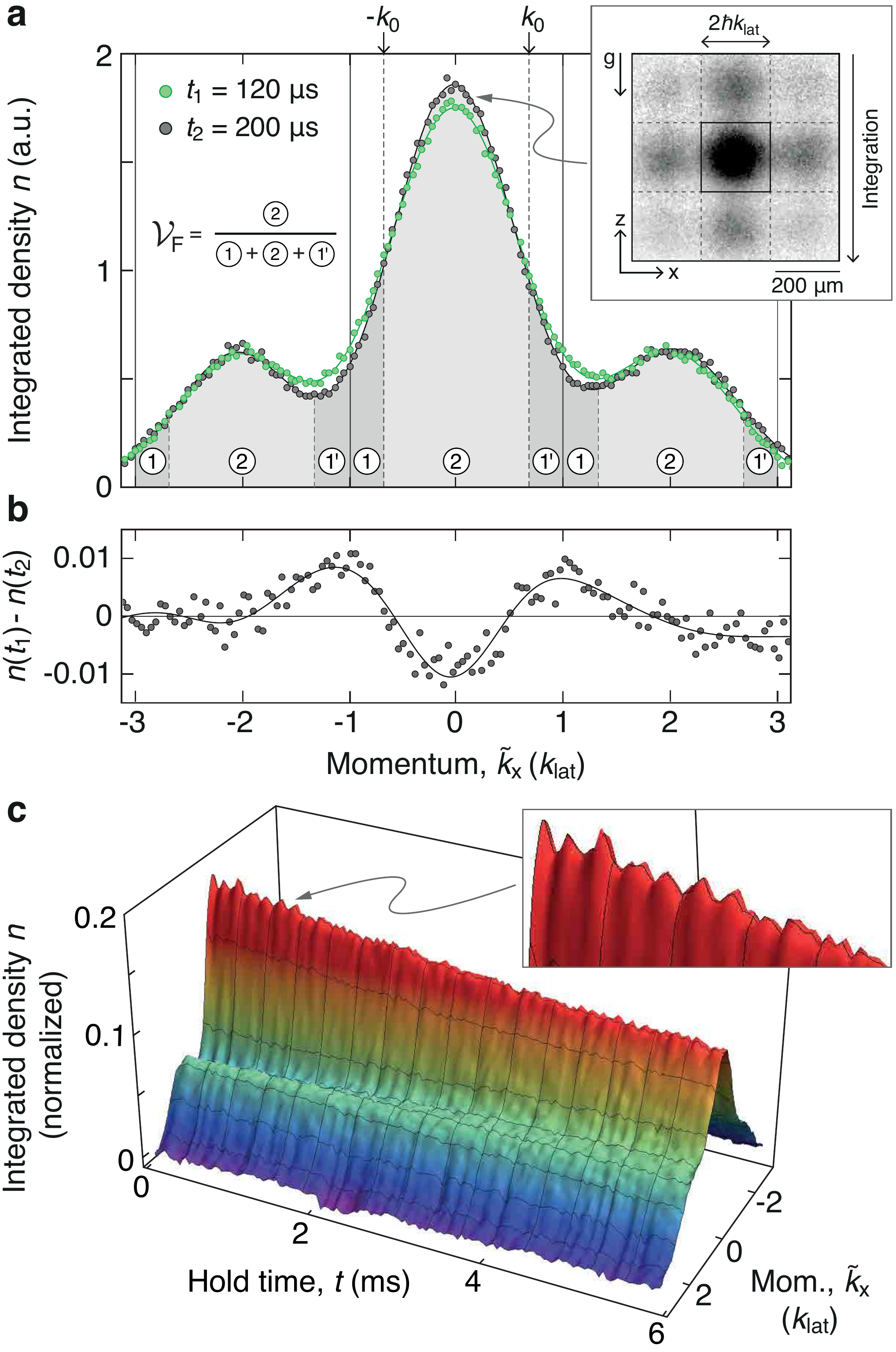}
  \end{center}
  \vspace*{-0.2cm}
  \caption{\label{fig:2} \textbf{Quench dynamics in the momentum
      distribution of fermi\-onic $^{40}$K atoms.} \textbf{a,} After
    variable hold times $t$, the fermionic momentum distributions are
    recorded after time-of-flight expansion (inset). Integration 
    along the z-direction results in one-dimensional momentum profiles 
    and averages the effect of Bloch oscillations, which occur as a 
    consequence of gravity. We show profiles at 
    an interspecies scattering 
    length of $a_{\FB} = -161.2(1)$ $a_0$ for $t_{1} = 120$ $\mu$s (green) 
    and $t_{2} = 200$ $\mu$s (black). Solid lines show a fit of three
    Gaussians that are separated by the Brillouin zone width $2 
    k_{\rm lat}$. \textbf{b,} The residual profile $n(\tilde{k}_x, t_{1}) -
    n(\tilde{k}_x, t_{2})$ illustrates the redistribution of population within
    the Brillouin zone. \textbf{c,} Evolution of the fermionic
    momentum distribution after the quench. The integrated profiles
    for each time $t$, $n(\tilde{k}_x, t)$, are normalised to a total area of
    one. Coherent quench dynamics is visible as a periodic modulation
    of the peak height at $\tilde{k}_x=0$ (zoomed in the inset).}
\end{figure}

The experiment begins with the preparation of a quantum degenerate
mixture of $2.1(4) \times 10^5$ fermionic $^{40}$K and $1.7(3) \times
10^5$ bosonic $^{87}$Rb atoms in their absolute hyperfine ground
states $| 9/2 , -9/2 \rangle $ and $| 1 , +1 \rangle$,
respectively. The temperature of the spin-polarised Fermi gas is
typically $T/T_{\rm F} = 0.20(2)$. Its interaction with the bosons is
tuned by means of a Feshbach resonance at $546.75(6)$
G\cite{Simoni08}, addressing interspecies scattering lengths $a_{\FB}$
in a range between $-161.2(1)$ $a_0$ and $+59(10)$
$a_0$. Subsequently, a 3D optical lattice operating at a wavelength of
$\lambda_{\rm lat} = 738$ nm is adiabatically ramped up within $50$ ms
to a depth of $V_{\rm L} = 3.5(2) E_{\rm rec}^{\rm F}$, where $E_{\rm
  rec}^{\rm F}= \hbar^2 k_{\rm lat}^2/(2 m_{\F})$ denotes the recoil
energy, $m_{\rm F}$  the atomic mass of $^{40}$K, and 
$k_{\rm lat}=2\pi/\lambda_{\rm lat}$ (for the corresponding lattice 
depths for $^{87}$Rb, see Methods). For these parameters, the fermions 
form a metallic many-body state within the first lattice band 
(see Methods)\cite{schneider08} and the bosons form a BEC. Then, we quench 
the system by rapidly increasing the lattice depth to $V_{\rm H}=18(1)
E_{\rm rec}^{\rm F}$, suppressing the tunnel coupling between
lattice sites and initiating coherent nonequilibrium dynamics of the
Fermi-Bose many-body state. After letting the system evolve for
variable hold times $t$, all trapping potentials are suddenly switched
off and an absorption image of the momentum distribution of $^{40}$K
is recorded after 9 ms time-of-flight expansion (see
Fig.~\ref{fig:2}{a}, inset).

The dynamics of the fermionic many-body state is revealed via
oscillations in the momentum distribution (see Fig.~\ref{fig:2}). The
recorded absorption images are integrated along the direction of
gravity to obtain 1D momentum profiles $n(\tilde{k}_x,t)$ at discrete hold
times $t$, sampled in steps of $40$ $\mu$s. Figure \ref{fig:2}{a}
compares two such profiles that are recorded at the approximate times
of a half and a full oscillation cycle for a fixed value of Fermi-Bose
interactions. The residual after subtracting the two profiles from
each other (see Fig.~\ref{fig:2}{b}) illustrates how the
interaction-driven dynamics leads to a redistribution of population
from momenta at the centre to momenta at the edge of the Brillouin
zone.  The coherent quench dynamics can be observed as a periodic
modulation of the peak height at $\tilde{k}_x=0$ for hold times shorter than
the time scale set by residual tunnelling of the fermions. For longer
times, equilibration dominates and the momentum profiles relax towards a state with a more uniform
distribution across the Brillouin zone (Fig.~\ref{fig:2}{c}) (see Methods).


\subsection{Fermionic visibility.}

\begin{figure}
  \begin{center}
    \includegraphics[width=1\columnwidth]{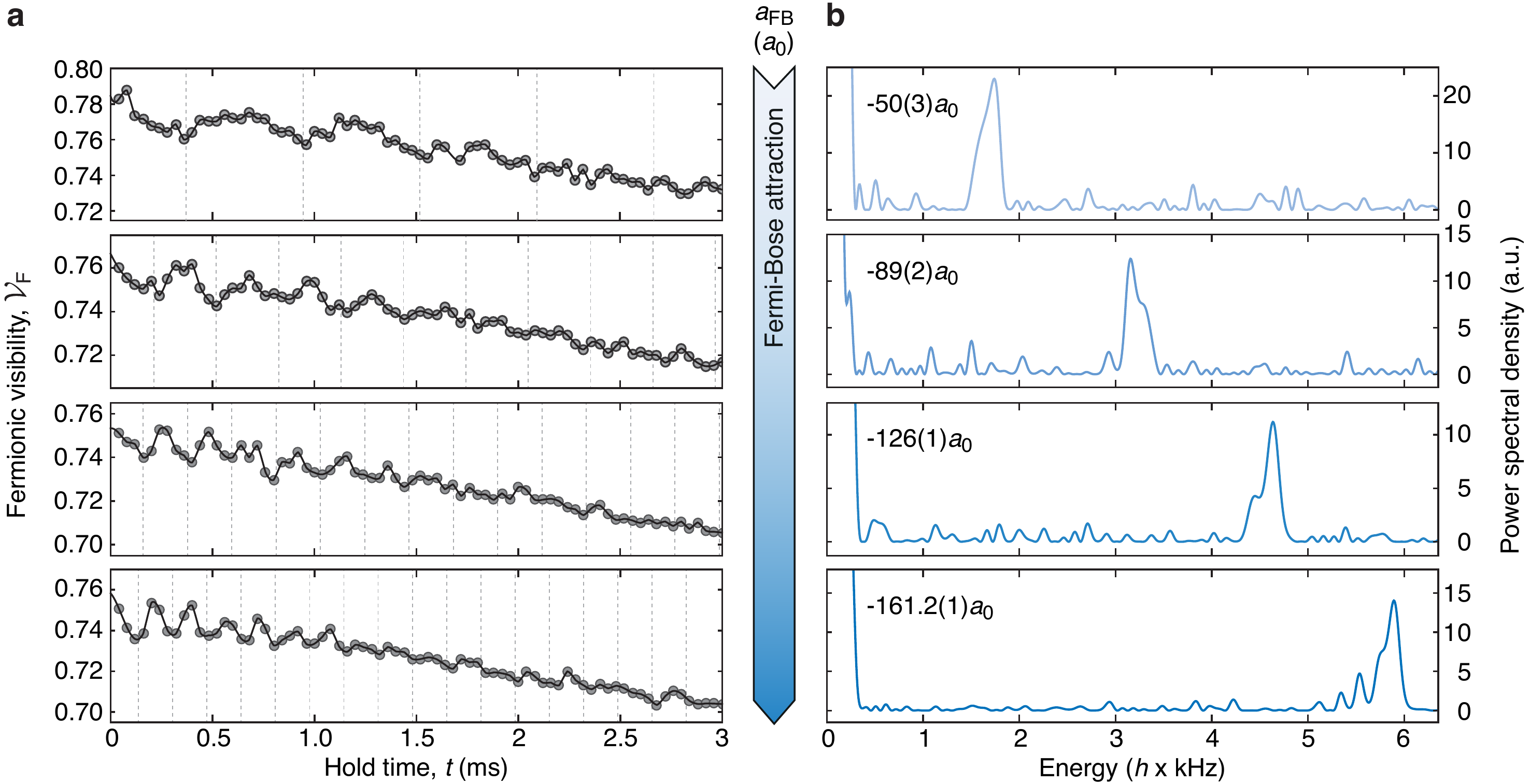}
  \end{center}
  \caption{\label{fig:3} \textbf{Coherent quench dynamics of a
      metallic many-body state.} \textbf{a,} Time traces of the
    visibility show the quantum dynamics of fermionic $^{40}$K atoms,
    driven by variable attractive interactions with bosonic $^{87}$Rb
    atoms. Dashed vertical lines indicate the periodicity of the
    dynamics. The solid lines interpolate the data and serve as a
    guide to the eye. Each data point corresponds to a single run of
    the experiment.  \textbf{b,} Fourier analysis reveals the spectral
    content of the time traces in (\textbf{a}). }
\end{figure}

A quantitative understanding of the coherent fermionic dynamics can be
gained from the time traces of the visibility $\mathcal{V}_{\rm F}$. For each 
momentum profile $n(\tilde{k}_x,t)$, the fermionic visibility is calculated 
after adding the regions with momenta $[-k_{\rm lat}, -3 k_{\rm lat}]$ and 
$[k_{\rm lat}, 3 k_{\rm lat}]$ to the first Brillouin zone 
$[-k_{\rm lat}, k_{\rm lat}]$, as illustrated in Fig.~\ref{fig:2}{a}. 
In order to obtain the largest amplitude of the visibility oscillations, 
we choose $k_0 = 2 k_{\rm lat} /3$, which is approximately the
$k$-value where the residual profile in Fig.~\ref{fig:2}{b}
changes sign.  Figure~\ref{fig:3}{a} shows time traces of the
evolution with up to ten observable oscillation periods. Upon
increasing the attraction between fermions and bosons, the period of
the oscillations becomes shorter as expected from the theoretical
analysis. This confirms that the quench dynamics is driven by the
interspecies interaction $U^{\rm FB} \propto a_{\rm FB}$. In general,
we observe oscillation amplitudes that are significantly smaller than
in the case of bosonic collapse and revival
dynamics\cite{greiner02,will10}. The reason is that only correlations 
between fermions on sites with bosons contribute to the dynamics, 
while unaccompanied fermions form a static background. The Fermi-Bose overlap 
volume is fundamentally limited because the in-trap size of a Fermi gas 
is significantly larger than a BEC with comparable atom numbers due to Pauli pressure. Additionally, 
differential gravitational sag can lead to a vertical displacement of the two 
atom clouds. In our setup, those two effects result in about 5\% of the 
fermions overlapping with bosons. This is compatible 
with the measured oscillation amplitudes (see Methods). Furthermore, 
finite temperature and Fermi-Bose interactions in the initial state 
localise fermions\cite{kohl05, best09} and reduce the visibility. 
Finally, residual tunnelling after the quench is expected to induce 
damping and to reduce the oscillation
amplitude\cite{fischer_schutzhold_08,wolf_hen_10_51}.


\newpage

\section*{Discussion}

The spectral content of the fermionic quench dynamics is revealed via
Fourier transform of the visibility time traces (see Methods).  As shown in Fig.~\ref{fig:3}{b}, the spectra are
dominated by a single peak, in remarkable contrast to the complex
spectra of the bosonic collapse and revival dynamics in the same
experimental setting\cite{will11}. Its width of about 300 Hz is
compatible with dephasing as a result of both residual tunnelling and
a small harmonic anticonfinement (see Methods). The peak also displays
a comb-like substructure with several frequencies of order $U^{\rm
  FB}/h$.  We assign this substructure to the deformation of on-site
orbitals as a result of interactions\cite{johnson09, best09, will10}
that effectively gives rise to an explicit dependence of the
Fermi-Bose interaction energy on the bosonic on-site occupation $n$,
$U^{\rm FB}_n$ (see Methods). According to Eq.~\eqref{eq:vis-notrap}
additional peaks at frequencies $2U^{\FB}$, $3U^{\FB}$, $\dots$ are
expected, but not observed in the spectra. As follows from our
discussion of the two-site system, such higher frequency components
result from correlations between fermions in lattice sites whose
occupations differ by two or more bosons. However, due to the
different sizes of the fermion and boson clouds, as well as their
differential gravitational sag (see Methods, Supplementary Figure 1 and Supplementary Note 1), such correlations are
strongly suppressed in our setup. This results in the suppression of
higher harmonics of $U^{\rm FB}$ below the noise level of our spectra.

\begin{figure}
  \begin{center}
    \includegraphics[width=0.55\columnwidth]{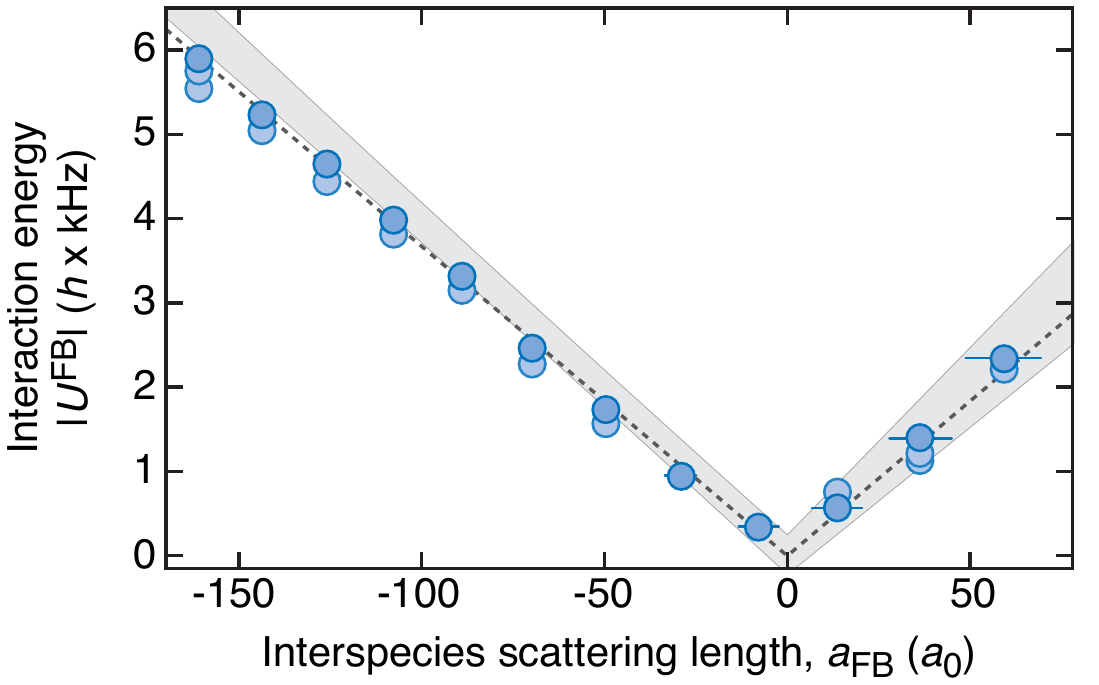}
  \end{center}
  \caption{\label{fig:4} \textbf{Precision measurement of the
      Fermi-Bose interaction energy $U^{\rm FB}$.} For each
    interspecies scattering length $a_{\rm FB}$ the dominant spectral
    components (dark points) and the spectral substructure (light
    points) are shown. Horizontal error bars reflect the experimental
    uncertainty (s.~d.) of the scattering length $a_{\rm FB}$,
    vertical error bars are smaller than the size of the data
    points. The dashed line shows a linear fit $C |a_{\rm FB}|$ to the
    dominant spectral features, yielding $C = 36.7(3)$ Hz$/a_0$. The
    shaded area shows a numerical calculation assuming the
    prescription $U^{\rm FB} = (2 \pi \hbar^2 a_{\rm FB}/\mu)\int dr^3
    |w_{\rm B}({\bf r})|^2 |w_{\rm F}({\bf r})|^2$ of a single-band
    Hubbard model, taking into account the experimental uncertainties
    in lattice depth and interspecies scattering length $a_{\rm
      FB}$. Here, $w_{\rm F}({\bf r})$ and $w_{\rm B}({\bf r})$ denote
    the fermionic and bosonic Wannier functions, respectively, and
    $\mu = m_{\rm F} m_{\rm B}/(m_{\rm F} + m_{\rm B})$ is the reduced
    mass.}
\end{figure}

In Fig.~\ref{fig:4}, we show the progression of $U^{\rm FB}$ as a
function of the interspecies scattering length both for attractive and
repulsive Fermi-Bose interactions. We compare the experimental results
to numerical calculations of $U^{\rm FB}$ that use Wannier functions
as on-site orbitals. The agreement is remarkable. On the attractive
side, the results of the calculations are compatible with the highest
frequency components measured experimentally. On the repulsive side,
all frequency components measured in the experiments are contained
within the bounds of the calculations.

\newpage


\newpage

In summary, we have observed coherent quench dynamics in metallic
states of ultracold fermionic atoms in an optical lattice.  In the
hybrid Fermi-Bose system investigated here, the time evolution arises
from the delocalised character of the initial fermionic state,
interspecies interactions, and an initial bosonic state that exhibits
site-to-site fluctuations of the atom number. Such coherent dynamics
also occurs in spin-1/2 interacting fermionic
systems\cite{iyer14,mahmud14}, and is expected to emerge in
higher-spin fermionic systems\cite{krauser12}, following a similar
quench protocol. The amplitude of the visibility oscillations depends
on the single-particle correlations between lattice sites.  Therefore,
coherent quench dynamics can serve as a novel tool to probe
correlations in delocalised quantum phases of fermionic systems, such
as the Hubbard model\cite{moeckel08} and chains of spin-polarised
fermions with intersite interactions\cite{manmana07}. This information
is complementary to the site-resolved precision measurements of
occupations in quantum gas
microscopes\cite{bakr_gillen_09,sherson_weitenberg_10}. Finally, the
spectral analysis of the visibility oscillations enables precision
measurements of on-site interactions and may be used to reveal complex
interaction effects in hybrid quantum many-body systems.

\begin{methods}

  \textbf{Experimental state preparation.} Fermi gases of
  $2.1(4)\times 10^5$ $^{40}$K atoms at a temperature of $T/T_{\rm F}
  = 0.20(2)$ and BECs of $1.7(3) \times 10^5$ $^{87}$Rb atoms were
  simultaneously created in the hyperfine states $|9/2,-9/2\rangle$
  and $|1,+1\rangle$, respectively. The degenerate Fermi-Bose mixtures
  were held in a pancake-shaped optical dipole trap operating at
  $\lambda_{\rm dip} = 1,030$ nm.  The interspecies scattering length
  $a_{\rm FB}$ between fermions and bosons was tuned by means of a
  Feshbach resonance, located at a magnetic field of $546.75(6)$
  G\cite{Simoni08}. The 3D optical lattice ($\lambda_{\rm lat} =
  738$nm) was operated at blue detuning with respect to the relevant
  atomic transitions of both $^{40}$K and $^{87}$Rb. It was
  adiabatically ramped to a depth of $V_{\rm L} = 3.5(2) E_{\rm
    rec}^{\rm F}$ for $^{40}$K [corresponding to $5.2(3) E_{\rm
    rec}^{\rm B}$ for $^{87}$Rb] within $50$ ms. The trapping
  frequencies of the horizontal and vertical confinement
  ($\omega_\perp$, $\omega_z$) were $2 \pi \times (36,173)$ Hz for
  $^{40}$K and $2 \pi \times (25, 94)$ Hz for $^{87}$Rb. Then, a
  non-adiabatic jump into a deep lattice, $V_{\rm H} = 18(1) E_{\rm
    rec}^{\rm F}$ for $^{40}$K [corresponding to $27(2) E_{\rm
    rec}^{\rm B}$ for $^{87}$Rb], was performed within $50$ $\mu$s,
  slow enough to avoid population of higher lattice bands, but fast
  with respect to tunnelling in the first band. Simultaneously with
  the lattice jump, the harmonic confinement in the horizontal plane
  was reduced to $- 2 \pi \times 16$ Hz for $^{40}$K and $2 \pi \times
  0$ Hz for $^{87}$Rb, enhancing the coherence time of the quench
  dynamics\cite{will10}. In the deep lattice, the tunnelling matrix
  elements for fermions and bosons are $J_{\rm F} = 2\pi \times 33$ Hz
  and $J_{\rm B} = 2\pi \times 3$ Hz, respectively. The corresponding
  tunnelling time scales are $\tau_{\rm F} = h/(z J_{\rm F}) = 5.1$ ms
  and $\tau_{\rm B} = h/(z J_{\rm B}) = 56$ ms, where $z=6$ is the
  coordination number of a 3D lattice.
    
  For the above loading parameters, the fermions form a metallic state
  with trap-averaged filling per lattice site of about $\bar{m} =
  k_{\rm F}/ k_{\rm lat} = 0.1$ for vanishing Fermi-Bose interactions
  ($a_{\rm FB} \sim 0$) and about $\bar{m} = 0.3$ for attractive
  Fermi-Bose interactions ($a_{\rm FB} \sim -125$ $a_0$)\cite{will11}.
  Accordingly, the fermionic momentum distributions recorded after $9$
  ms time-of-flight expansion display a partially filled first
  Brillouin zone (see Fig.~\ref{fig:2}). The bosons form a BEC with a
  trap-averaged filling per lattice site of about $\bar{n} =
  |\alpha|^2 = 1$ and a maximal filling in the trap centre of $2.5$
  atoms.

  \noindent \textbf{In-trap arrangement of atomic clouds.} For the
  above loading parameters the horizontal and vertical in-situ
  Thomas-Fermi radii $(r_\perp, r_z)$ are about (50 $\mu$m, 11 $\mu$m)
  for $^{40}$K and (21 $\mu$m, 5 $\mu$m) for $^{87}$Rb. Although the
  total atom numbers are comparable, the fermionic cloud is about 10
  times larger in volume than the bosonic one, as a consequence of
  Pauli pressure. The differential gravitational sag between the
  clouds has been measured to be 8(2) $\mu$m, leading to a notable
  displacement (see Supplementary Figure 1 and Supplementary Note 1). Only the overlap
  volume of fermions and bosons (plus a thin shell of few lattice
  sites, which represents the coherence length of the fermions)
  contributes to the fermionic quench dynamics; about 5\% of the
  fermions overlap with the bosons. This is compatible with the
  amplitude of the fermionic quench dynamics shown in
  Fig.~\ref{fig:2}{c}, corresponding to about 5\% of the atomic
  density in momentum space.

\noindent \textbf{Spectral analysis.} The visibility time traces of the quench dynamics typically cover an
observation time of $6$ ms, sampled in steps of $40$ $\mu$s. In order
to obtain high-resolution, low-noise spectra, the time traces are
processed as follows: The raw data points are interpolated using cubic
splines. The origin of the time axis, $t=0$, corresponds to the
beginning of the jump from $V_{\rm L}$ to $V_{\rm H}$.  In order to
avoid distortion of the spectral analysis due to dynamics that slowly
starts during the jump, the first $40$ $\mu$s of the interpolated
trace are removed.  For times longer than the observation time, we
smoothly attach an exponential decay with a time scale of about $2$ ms
to the interpolated curve. The such prepared curve is concatenated to
its mirror image, which is obtained upon exchanging time $t$ by
$-t$. The resulting trace is again sampled in steps of $40$ $\mu$s and
numerical Fourier analysis is performed. The processing scheme
improves the data quality in two ways: First, the knowledge of the
initial phase allows to mirror the data. This doubles the size of the
data set and yields a two-fold improvement of the spectral resolution
to about $85$ Hz. Second, the additional extension of the data set by
a smooth exponential decay avoids high frequency artefacts, which
would arise from Fourier transform of sharp cut-offs, and makes the
Fourier spectra quasi-continuous.

  \noindent \textbf{Outline of the calculation.}  We outline the
  derivation of Eq.~\eqref{eq:vis-notrap} and discuss its extension to 3D.
The Hamiltonian governing the time evolution after the quench is
  given by
  \begin{equation}\label{eq:onsiteham}
    \hat{H} = \sum_{j\in M} \left[
      \frac{U^{\BB}}{2}\hat{a}^{\dag}_{j}\hat{a}^{}_{j}(\hat{a}^{\dag}_{j}\hat{a}^{}_{j}-1) + 
      U^{\FB}\hat{a}^{\dag}_{j}\hat{a}^{}_{j}\hat{c}^\dag_{j}\hat{c}^{}_{j}\right],
  \end{equation}
  $\hat{a}^{\dag}$ and $\hat{c}^\dag$ being bosonic and fermionic
  creation operators respectively and $M$ is the number of lattice
  sites.

  The explicit action of the time evolution operator on the initial
  state $\ket{\psi_{0}} = \ket{\psi_{\FS}}\otimes\ket{\psi_{\rm BEC}}$
  is (setting $\hbar=1$ for convenience)
  \begin{equation}
    \label{eq:psi(t)-notrap}
    \begin{split}
      &\ket{\psi(t)} = e^{-i\hat{H}t}\ket{\psi_{0}} \\
      &=
      e^{-M\frac{|\alpha|^{2}}{2}}M^{-\frac{N}{2}}\sum_{\{n_{\mathbf{s}}\}}\sum_{\{{\bf
          r}_{j}\in M\}}
      \left[\prod_{j=1}^{N}e^{i\mathbf{k}_{j}\cdot\mathbf{r}_{j}}\right]
      \left[\prod_{\mathbf{s}\in M}
        \frac{\alpha_{\mathbf{s}}^{n_{\mathbf{s}}}}{n_{\mathbf{s}}!}\right]
      e^{-it\frac{U^{\BB}}{2}\sum_{\mathbf{l}\in M}\hat{a}^{\dag}_{\bf
          l}\hat{a}_{\bf l}(\hat{a}^{\dag}_{\bf l}\hat{a}_{\bf l}-1)}
      e^{-it U^{\FB}\sum_{\mathbf{l}\in M}\hat{a}^{\dag}_{\bf l}\hat{a}_{\bf l} \hat{c}^{\dag}_{\bf l}\hat{c}_{\bf l}} \\
      &\hspace{0.71\textwidth} \times \prod_{\mathbf{s}\in
        M}(\hat{a}^{\dag}_{\mathbf{s}})^{n_{\mathbf{s}}}\prod_{j=1}^{N}\hat{c}^\dag_{\mathbf{r}_{j}}\ket{0}\\
      &=
      e^{-M\frac{|\alpha|^{2}}{2}}M^{-\frac{N}{2}}\sum_{\{n_{\mathbf{s}}\}}\sum_{\{{\bf
          r}_{j}\in M\}}
      \left[\prod_{j=1}^{N}e^{i\mathbf{k}_{j}\cdot\mathbf{r}_{j}}\right]
      \left[\prod_{\mathbf{s}\in M}
        \frac{\alpha_{\mathbf{s}}^{n_{\mathbf{s}}}}{n_{\mathbf{s}}!}
        e^{-it\frac{U^{\BB}}{2}n_{\mathbf{s}}(n_{\mathbf{s}}-1)}
      \right] e^{-it U^{\FB}\sum_{\mathbf{l}\in M}\sum_{i=1}^{N}
        n_{\mathbf{l}}\delta_{\mathbf{l}\mathbf{r}_{i}}} \\
      &\hspace{0.71\textwidth} \times \prod_{\mathbf{s}\in
        M}(\hat{a}^{\dag}_{\mathbf{s}})^{n_{\mathbf{s}}}\prod_{j=1}^{N}\hat{c}^\dag_{\mathbf{r}_{j}}\ket{0}\\
      & =
      e^{-M\frac{|\alpha|^{2}}{2}}M^{-\frac{N}{2}}\sum_{\{n_{\mathbf{s}}\}}\sum_{\{{\bf
          r}_{j}\in M\}}
      \left[\prod_{j=1}^{N}e^{i\mathbf{k}_{j}\cdot\mathbf{r}_{j} - it
          U^{\FB} n_{\mathbf{r}_j}}\right] \left[\prod_{\mathbf{s}\in
          M}
        \frac{\alpha_{\mathbf{s}}^{n_{\mathbf{s}}}}{n_{\mathbf{s}}!}
        e^{-it\frac{U^{\BB}}{2}n_{\mathbf{s}}(n_{\mathbf{s}}-1)}
      \right] \prod_{\mathbf{s}\in
        M}(\hat{a}^{\dag}_{\mathbf{s}})^{n_{\mathbf{s}}}\prod_{j=1}^{N}\hat{c}^\dag_{\mathbf{r}_{j}}\ket{0}\\
      &\equiv
      e^{-M\frac{|\alpha|^{2}}{2}}M^{-\frac{N}{2}}\sum_{\{n_{\mathbf{s}}\}}\sum_{\{{\bf
          r}_{j}\in M\}} \Phi(\{n_{\mathbf{s}}\},\{\mathbf{r}_j\})
      \prod_{\mathbf{s}\in
        M}(\hat{a}^{\dag}_{\mathbf{s}})^{n_{\mathbf{s}}}\prod_{j=1}^{N}\hat{c}^\dag_{\mathbf{r}_{j}}\ket{0}.
    \end{split}
  \end{equation}
  We first evaluate the expectation value of the density matrix
  $\xv{\hat{c}^\dag_{\mathbf{m}}\hat{c}^{}_{\mathbf{n}}}$ as a step
  towards calculating the momentum distribution,
  \begin{equation}\label{eq:dmat-calc}
    \begin{split}
      &\xv{\hat{c}^\dag_{\mathbf{m}}\hat{c}^{}_{\mathbf{n}}} =
      \frac{e^{-M|\alpha|^{2}}}{M^{N}}
      \sum_{\{n'_{\mathbf{s}}\}}\sum_{\{{\bf r}'_{j}\in M\}}
      \sum_{\{n_{\mathbf{s}}\}}\sum_{\{{\bf r}_{j}\in M\}}
      \Phi^{*}(\{n'_{\mathbf{s}}\},\{\mathbf{r}'_j\})\Phi(\{n_{\mathbf{s}}\},\{\mathbf{r}_j\})\\
      &\hspace{0.4\textwidth} \times \bra{0}\prod_{\mathbf{s}\in
        M}(\hat{a}_{\mathbf{s}})^{n'_{\mathbf{s}}}\prod_{j=1}^{N}\hat{c}_{\mathbf{r}'_{j}}
      \hat{c}^\dag_{\mathbf{m}}\hat{c}_{\mathbf{n}}
      \prod_{\mathbf{s}\in
        M}(\hat{a}^{\dag}_{\mathbf{s}})^{n_{\mathbf{s}}}\prod_{j=1}^{N}\hat{c}^\dag_{\mathbf{r}_{j}}\ket{0}\\
      &= \frac{e^{-M|\alpha|^{2}}}{M^{N}}\sum_{\{{\bf r}'_{j}\in M\}}
      \sum_{\{n_{\mathbf{s}}\}}\sum_{\{{\bf r}_{j}\in M\}}
      \Phi^{*}(\{n_{\mathbf{s}}\},\{\mathbf{r}'_j\})\Phi(\{n_{\mathbf{s}}\},\{\mathbf{r}_j\})
      \left[\prod_{\mathbf{s}\in M}n_{\mathbf{s}}!\right] \bra{0}
      \prod_{j=1}^{N} \hat{c}_{\mathbf{r}'_{j}}
      \hat{c}^\dag_{\mathbf{m}}\hat{c}_{\mathbf{n}} \prod_{j=1}^{N}
      \hat{c}^\dag_{\mathbf{r}_{j}}\ket{0}.
    \end{split}
  \end{equation}
  The cases $\mathbf{m}\neq\mathbf{n}$ and $\mathbf{m}=\mathbf{n}$
  have to be treated separately, yielding
  \begin{equation}\label{eq:dmat}
    \bra{\psi(t)}\hat{c}^\dag_{\mathbf{m}}\hat{c}^{}_{\mathbf{n}}\ket{\psi(t)}
    = \frac{1}{M}\left\{(1-\delta_{\mathbf{m}\mathbf{n}})
      \left[ \sum_{j=1}^{N} e^{i \mathbf{k}_{j}\cdot (\mathbf{n}-\mathbf{m})}\right]
      e^{|\alpha_{\mathbf{n}}|^2 (e^{-it U^{\FB}}-1)}e^{|\alpha_{\mathbf{m}}|^2 (e^{it U^{\FB}}-1)} 
      + \delta_{\mathbf{m}\mathbf{n}} N\right\}.
  \end{equation}
  The sum in the brackets in Eq.~\eqref{eq:dmat} cannot be calculated
  analytically in 3D. This is due to the constraint that the fermions
  fill up the lowest energy states governed by $E_{\mathbf{k}}<E_{\F}$
  with $E_{\mathbf{k}}=-J[\cos(k_{x}a)+\cos(k_{y}a)+\cos(k_{z}a)]$,
  where $J$ is the hopping.  In one dimension, however, the sum is
  easily carried out. For unconfined bosons, the site occupation
  $|\alpha_{\mathbf{m}}|^{2}\equiv |\alpha|^{2}$ is a constant equal
  to the mean number of bosons per site. We compute the Fourier
  transform of Eq.~\eqref{eq:dmat} to get the momentum distribution,
  and integrate to obtain the visibility. This gives the expression in
  Eq.~\eqref{eq:vis-notrap} for the 1D visibility.

  In 3D, it is possible to obtain an analytical expression for small
  fermionic filling, where the Fermi surface is approximately a
  sphere. The visibility in this case is,
  \begin{equation}\label{eq:vis-3d-cont}
    {\cal V}_{F}(t) = \frac{k_{0}}{k_{\rm lat}} + 
    \left[\frac{k_{0}}{2k_{\F}}\left(3-\frac{k_{0}^{2}}{k_{\F}^{2}}\right)-
      \frac{k_{0}}{k_{\rm lat}}\right]
    e^{2|\alpha|^{2} [\cos(tU^{\FB})-1]}
  \end{equation}
  for $k_{0}<k_{\F}$.  
  
  \noindent\textbf{Effects of harmonic confinement.}  We assume that
  the single particle ground state of a harmonically trapped system in
  a lattice can be described by the ground state in the continuum with
  a lattice renormalised mass\cite{rigol_muramatsu_04}. It is not
  possible to analytically study a trapped lattice system. We further
  assume that all the bosons are in the ground state.  The average on-site
   occupancies then take the form
  $|\alpha_{\mathbf{m}}|^{2}=|\alpha|^{2}e^{-\nu_{\rm
      B}|\mathbf{m}|^{2}}$, where $|\alpha|^{2}$ denotes the average
  occupation in the centre of the trap and $1/\sqrt{\nu_{\rm B}}\equiv
  \sqrt{\frac{2\hbar}{m_{\rm B}\omega_{\rm B}}}$ is the length scale
  of the trap.  For this case, the visibility (in the limit of low
  fermionic filling) is given by
  \begin{equation}
    \label{eq:vis-bose-trap}
    {\cal V}_{F}(t) = \frac{k_{0}}{k_{\rm lat}} + 
    \left[\frac{k_{0}}{2k_{\F}}\left(3-\frac{k_{0}^{2}}{k_{\F}^{2}}\right)-
      \frac{k_{0}}{k_{\rm lat}}\right]
    \int{\rm d}^{3}{\bf m}\,
    e^{2|\alpha|^{2}e^{-\nu'_{\B}|\mathbf{m}|^{2}} [\cos(tU^{\FB})-1]},
  \end{equation}
  where $\nu_{\B}'\equiv \nu_{\B}L^{2}$ and $L$ is the length of the
  fermionic system. Since the coordinates are rescaled, the
  integration is carried out over a cube of length one centred at the
  origin.  The amplitude of oscillations in
  Eq.~\eqref{eq:vis-bose-trap} is governed by
  $\int_{\mathbf{m}}e^{-4|\alpha|^{2}e^{-\nu'_{\B}|\mathbf{m}|^{2}}}$.
  As $|\alpha|^{2}$ increases, the amplitude increases for fixed
  $\nu_{\B}'$.  As $\nu_{\B}'$ increases with $\alpha$ fixed, the
  bosonic wave function becomes sharply peaked in space, decreasing
  the overlap between the bosons and the fermions. This in turn
  decreases the oscillation amplitude. Confinement of the bosons is
  therefore one reason for the reduced amplitude of oscillations seen
  in the experimental data.

  If we further include a confining trap for the fermions, one can no
  longer use the plane waves for the initial state.  Instead, one must
  use fermionic harmonic oscillator states.  As for the trapped
  bosons, we carry out calculations in the continuum since the
  eigenstates of harmonically trapped fermions in a lattice are not
  known analytically\cite{rigol_muramatsu_04,rey_thesis2004}. The
  masses of the atoms are renormalised masses obtained from the
  low-density limit in the
  lattice\cite{rigol_muramatsu_04,rey_thesis2004}.  We get the
  following expression for the fermionic visibility:
  \begin{multline}
    \label{eq:vis-traps}
    {\cal V}_{\F}(t) = \sum_{l=1}^{N}\int {\rm d}x\,{\rm d}y\, {\rm
      d}z\, {\rm d}z'\,
    \Bigg[\psi_{\F,l}^{*}(x,y,z')\psi_{\F,l}(x,y,z)
    e^{|\alpha(x,y,z')|^{2}(e^{itU^{\FB}}-1)+
      |\alpha(x,y,z)|^{2}(e^{-itU^{\FB}}-1)}\\
    \times \frac{\sin(k_{0}|y-y'|)}{\pi N|y-y'|} \Bigg] +
    \frac{k_{0}}{k_{\rm lat} N}\sum_{l=1}^{N} \int {\rm d}x\,{\rm
      d}y\, {\rm d}z\, |\psi_{\F,l}(x,y,z)|^{2} \left[1-
      e^{2|\alpha(x,y,z)|^{2}\{\cos(tU^{\FB})-1\}} \right]
  \end{multline}
  where the $\psi_{F,l}$ are harmonic oscillator wave functions. The
  sum over $l$ is shorthand for the sum over the set of quantum
  numbers describing a harmonic oscillator as we fill states.  In the
  thermodynamic limit, as $N\to\infty$, the primary contribution to
  the first term in Eq.~\eqref{eq:vis-traps} comes from the diagonal
  part $z=z'$. In this limit, the integrand is proportional to
  $\delta(z-z')$, but the prefactor has to be determined
  numerically. With this, we get the compact expression
  \begin{equation}
    \label{eq:vis-trap-td}
    {\cal V}_{\F}(t) = \frac{k_{0}}{k_{\rm lat}} +
    \left({\cal V}_{0}-\frac{k_{0}}{k_{\rm lat}}
    \right)
    \int{\rm d}^{3}{\bf r}\, \rho_{\F}(\nu_{\F},r)e^{2|\alpha|^2
      e^{-\nu_{\B}r^{2}} \left(\cos U^{\FB}t -1\right)}
  \end{equation}
  where $\rho_{\F}$ is the density of harmonically trapped fermions,
  given by the Thomas-Fermi formula in the thermodynamic
  limit. $\nu_{\B,\F}$ are the inverse square length scales of the
  bosonic and fermionic traps respectively. ${\cal V}_{0}$ is
  calculated numerically from ${\cal V}_{\F}(t)$ at $t=0$.  The above
  expression assumes the thermodynamic limit. We have verified that
  calculations with experimental parameters exhibit negligible finite
  size effects, and the results agree with Eq.~\eqref{eq:vis-trap-td}.
  The non-uniform spatial distribution of fermions 
  contributes to a decrease in the oscillation amplitude. Differing
  confinement scales for the fermions and bosons affect the spatial
  overlap between them and additionally reduces the oscillation amplitude.  Damping of
  the oscillations in the experiment is dominantly due to residual
  tunnelling in the post quench system\cite{iyer14} and interactions
  between fermions and bosons in the initial state.  To emphasize a
  key point, in all the cases we have considered, the basic time dependence
  of the visibility oscillations remains the same.

  \noindent \textbf{Substructure of spectral features.}  The comb-like
  substructure of the peaks in Fig.~\ref{fig:3}{b} originates
  from occupation dependent interaction strengths $U^{\FB}_{n}$,
  corresponding to the interaction energy of a fermion and a boson on
  sites that contain one fermion and $n$ bosons\cite{will11}.
  Combining this modification with Eqs.~\ref{eq:dmat} and \ref{eq:vis-3d-cont}, the single-particle density matrix
  $\langle \hat{c}^{\dag}_{i}\hat{c}_{j}\rangle$ contains terms
  proportional to $\sum_{n_{i},n_{j}}\mathcal{C}\exp(-i
  U^{\FB}_{n_{i}} n_{i} t/\hbar+iU^{\FB}_{n_{j}}n_{j}t/\hbar)$, where
  $\mathcal{C}$ is independent of time.  Consequently, the spectrum
  contains the frequencies $(U^{\FB}_{n_{i}} n_{i} -
  U^{\FB}_{n_{j}}n_{j})/h$ for all integer values of $n_{i}$ and
  $n_{j}$, i.e.,~spectral features are expected at $0$,
  $U^{\FB}_{1}/h$, $(2U^{\FB}_{2}-U^{\FB}_{1})/h,\dots$.

\end{methods}

\newpage


\bibliographystyle{naturemag}


\begin{addendum}
  
\item We are indebted to Immanuel Bloch for generous support of the
  experimental efforts and advice during the preparation of the
  manuscript. We acknowledge Thorsten Best and Simon Braun for
  experimental assistance, and Ulf Bissbort and David Weiss for
  critical reading of the manuscript. This work was supported by the
  Deutsche Forschungsgemeinschaft (S.W.), the US Army Research Office
  with funding from the Defense Advanced Research Projects Agency
  (Optical Lattice Emulator program) (S.W.), the Office of Naval
  Research (D.I. and M.R.), the Graduate School Materials Science in
  Mainz (S.W.), and the Gutenberg-Akademie (S.W.).
\item[Author Contributions] S.W. conceived the experiment, carried out
  the measurements and analysed the data. D.I. and M.R. developed the
  theoretical model. All authors contributed significantly to the
  writing of the manuscript.
\item[Additional Information] Supplementary Information is available
  in the online version of the paper. Reprints and permissions
  information is available at www.nature.com/reprints. Correspondence
  and requests for materials should be addressed to S.W.~(email:
  sewill@mit.edu).
\item[Competing Interests] The authors declare no competing financial
  interests.
\end{addendum}

\newpage

\beginsupplement

\section*{Supplementary Information}

\section*{Note 1 - Additional information on the in-trap arrangement of atomic clouds}

For the loading parameters of the shallow lattice $V_{\rm L}$, 
the horizontal and vertical in-situ Thomas-Fermi radii $(r_\perp, r_z)$ are about 
(50 $\mu$m, 11 $\mu$m) for $^{40}$K and (21 $\mu$m, 5 $\mu$m) for $^{87}$Rb. Although the 
total atom numbers are comparable, the fermionic cloud is about 10 times larger in 
volume than the bosonic one. This is a consequence of Pauli pressure. The differential 
gravitational sag between the clouds has been measured to be 8(2) $\mu$m. This value 
was confirmed by a detailed mathematical model of the optical dipole trap in use.\cite{will11:2} 
Following the above assessment, the in-trap arrangement of the ellipsoidal fermion 
and boson clouds is illustrated to scale in Supplementary Figure~\ref{fig:1}. 

Only the volume in which fermions and bosons overlap (plus a thin shell of few lattice 
sites, which represents the coherence length of the fermions) contributes to the 
fermionic quench dynamics. For our parameters about 5\% of the fermions do overlap 
with the bosons. This is compatible with the amplitude of the fermionic quench dynamics 
shown in Fig.~2c of the main text, corresponding to about 5\% of the atomic density in momentum space.

In addition to the oscillation amplitude of the observed quench dynamics, the in-trap arrangement 
of the Fermi-Bose mixture explains the absence of higher harmonics $n \cdot U^{\rm FB}$ from the 
spectra. For example, a peak at $2 U^{\rm FB}$ appears for correlations between lattice sites that have 
a difference in boson number of $\Delta n = 2$ (see two-site model in the main text). This requires 
that fermions have overlap with the high density region of the boson cloud. However, for our 
experimental parameters (see Methods) a significant bosonic occupation of two can only be found 
in the centre of the bosonic cloud, where in turn the fermionic density - since it is the rim of 
the fermion cloud - is already much smaller than the mean value of the density of about 0.1 fermion 
per lattice site. Therefore, correlations between lattice sites that would give rise to higher 
harmonics are suppressed below the 10\% noise level that is present in the experimental spectra 
(compare Fig.~3 of the main text). Three-body loss could be an additional explanation for the 
absence of higher harmonics. However, based on previous observations in our lattice setup\cite{best09} 
enhanced three-body loss of sites with one fermion and $n>1$ bosons should not
play a significant role on the observation time scales and parameters of the present experiment.

\begin{figure}
  \begin{center}
    \includegraphics[width=0.8\columnwidth]{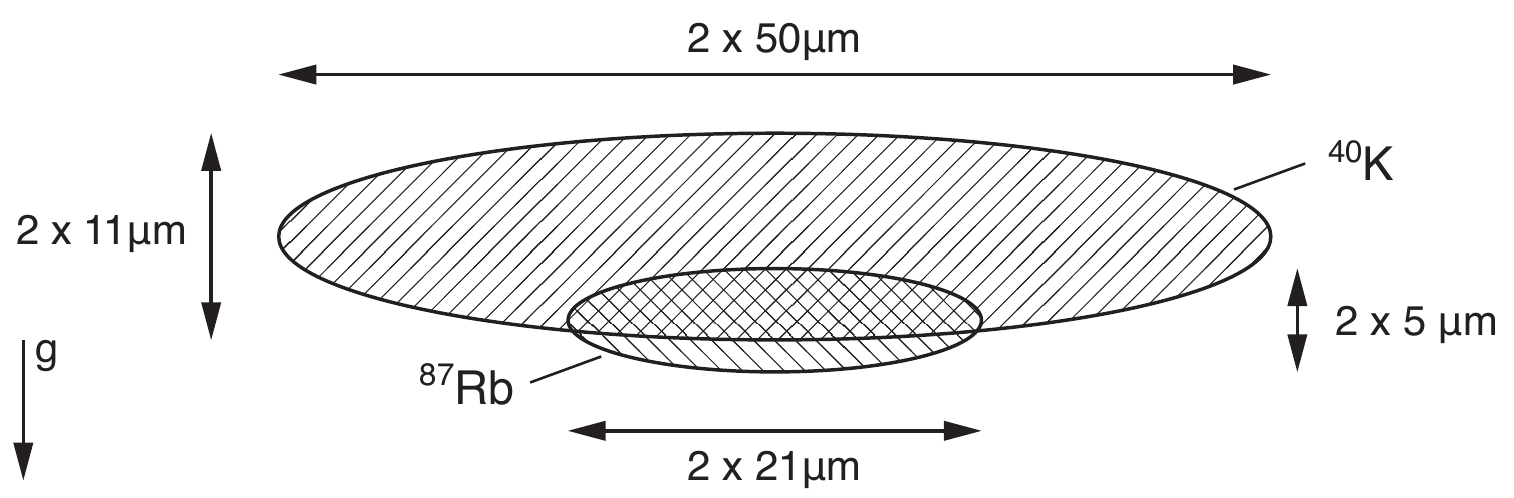}
  \end{center}
  \caption{\label{fig:1} In-trap cloud sizes and relative positions (to scale) of $^{87}$Rb and $^{40}$K  
  in the shallow lattice $V_{\rm L}$ for the experimental parameters. The centre of the Rb cloud 
  is located $8$ $\mu$m below the centre of the K cloud due to gravitational sag. 
  The direction of gravity is indicated on the left.}
\end{figure} 

\section*{Supplementary References}

\bibliographystyle{naturemag}

\end{document}